\documentclass[footinbib, twocolumn,english, aps, prb, showpacs, superscriptaddress, floats, amsmath, amssymb, floatfix]{revtex4}

\usepackage{color}
\usepackage{amsmath}
\usepackage{graphicx}
\usepackage{graphics}
\usepackage{pdftexcmds}
\usepackage{amssymb}
\usepackage{esint}
\usepackage{comment}
\usepackage{physics}
\usepackage{bm}

\newcommand{\Sp}{\mathbf{S}}
\newcommand{\tq}{{\ensuremath{\frac{3}{4}}}}
\def \bea {\begin{eqnarray}}
\def \eea {\end{eqnarray}}

\allowdisplaybreaks

\begin{document}

\title{Anisotropy driven response of skyrmion lattice in MnSc$_2$S$_4$ to applied magnetic fields}

\author{H. D. Rosales}
\affiliation{Instituto de F\'isica de L\'iquidos y Sistemas Biol\'ogicos (IFLYSIB), UNLP-CONICET, Facultad de Ciencias Exactas, La Plata, Argentina}
\affiliation{Departamento de F\'isica, Facultad de Ciencias Exactas, Universidad Nacional de La Plata, La Plata, Argentina}
\affiliation{Departamento de Cs. B\'asicas, Facultad de Ingenier\'ia, Universidad Nacional de La Plata, C.C. 67, 1900 La Plata, Argentina}

\author{F. A. G\'omez Albarrac\'in}
\affiliation{Instituto de F\'isica de L\'iquidos y Sistemas Biol\'ogicos (IFLYSIB), UNLP-CONICET, Facultad de Ciencias Exactas, La Plata, Argentina}
\affiliation{Departamento de F\'isica, Facultad de Ciencias Exactas, Universidad Nacional de La Plata, La Plata, Argentina}
\affiliation{Departamento de Cs. B\'asicas, Facultad de Ingenier\'ia, Universidad Nacional de La Plata, C.C. 67, 1900 La Plata, Argentina}

\author{K. Guratinder}
\affiliation{Laboratory for Neutron Scattering and Imaging, Paul Scherrer Institute, CH-5232 Villigen, Switzerland}
\affiliation{Department of Quantum Matter Physics, University of Geneva, CH-1211 Geneva, Switzerland }

\author{V. Tsurkan}
\affiliation{Experimental Physics V, University of Augsburg, D-86135 Augsburg, Germany}
\affiliation{Institute of Applied Physics, str. Academiei 5, MD 2028, Chisinau, Republic of Moldova}

\author{L. Prodan}
\affiliation{Experimental Physics V, University of Augsburg, D-86135 Augsburg, Germany}
\affiliation{Institute of Applied Physics, str. Academiei 5, MD 2028, Chisinau, Republic of Moldova}

\author{E. Ressouche}
\affiliation{Univetsit\'e Grenoble Alpes, CEA, INAC-MEM, Grenoble, France}

\author{O. Zaharko}\email{oksana.zaharko@psi.ch}
\affiliation{Laboratory for Neutron Scattering and Imaging, Paul Scherrer Institute, CH-5232 Villigen, Switzerland}

\date{\today}

\begin{abstract}
We theoretically and experimentally study the stability of the unconventional fractional antiferromagnetic skyrmion lattice (AF-SkL) in Mn$_2$S$_4$ spinel under magnetic fields applied along the $[$1-10$]$ crystal direction. By performing numerical Monte Carlo simulations for the minimal effective spin model that we  proposed in Ref. [S. Gao, et al.,  Nature 586, 37-41 (2020)], we show that the skyrmion lattice is aligned within the equivalent and symmetric $[$1-11$]$ or $[$1-11$]$ planes, which are equally inclined to the applied magnetic field. We attribute this behavior to the magnetic anisotropy of the host material. Neutron single crystal diffraction presents a very good agreement with the predictions of the effective model. It reveals that the topological spin texture gets destabilized at low temperatures and moderate magnetic fields and is replaced by a conical phase for B// $[$1-10$]$.  The present study elucidates the central role of the magnetic anisotropy in the stabilization of antiferromagnetic skyrmionic states.

\end{abstract}

\maketitle

\section{Introduction}{\label{Sec1}}

The notion of skyrmions, first introduced in particle physics to account for stability of quantized topological detects\cite{Skyrme1961}, is used broadly now also in magnetic systems. The prospect to create, move and annihilate isolated nanometer-sized swirls of spins is attractive for future memory devices\cite{Fert2013}. The bulk counterparts, so called skyrmion lattices (SkL) or skyrmion crystals\cite{Tokura2021}, are appropriate model systems to explore the microscopic mechanisms of formation of such magnetic quasiparticles, their topological properties and responses to external incites\cite{Nagaosa2013}.\\
Several microscopic mechanisms stabilising SkL are presently identified. The most established one enacts in noncentrosymmetric crystals through the asymmetric Dzyaloshinskii-Moriya (DM) interaction\cite{Bogdanov1989, Muelbauer2009}. The famous example is the B20-family, which consists of soft chiral ferromagnets (such as metallic MnSi\cite{Muelbauer2009}, isolating Cu$_2$OSeO$_3$\cite{Seki2012, White2014}) with Bloch-type SkL following the magnetic field. Another noncentrosymmetric host is
the polar lacunar spinel GaV$_4$X$_8$ with X= S, Se\cite{Kezsmarki2015}, where the Neel-type SkL are locked to the polar axis.
However, SkL are found also in centrosymmetric crystals. For example, in the intermetallic compound Gd$_2$PdSi$_3$ the localised 4f-moments on the triangular lattice are frustrated yielding a topologically nontrivial triple-$\bf{q}$ state\cite{Kurumaji2019, Okumura2020}.\\
Materials hosting SkL bear predominantly ferromagnetic interactions and their temperature-magnetic field (T, B) phase diagrams share a number of common features (see Fig.2 of Ref.\onlinecite{Seki2012} for a typical example). The zero field ground state is usually a helical structure, which in applied magnetic field attains a ferromagnetic component and transforms first into a conical phase and then into a field-polarized ferromagnet. The skyrmion phase is commonly located in an intermediate T-B range, adjacent to the conical phase from the low-T side and bordering to the paramagnetic state at the high-T side. The region of the SkL existence varies significantly, ranging from a tiny 'pocket' in MnSi\cite{Muelbauer2009} to an extensive area in Co-Zn-Mn alloys.\cite{Karube2016} It becomes increasingly recognised that magnetocrystalline anisotropy plays a significant role in the formation and stability of SkL's\cite{Preissinger2021}.\\
Materials with predominantly antiferromagnetic interactions might also host skyrmion lattices (AF-SkL)\cite{Okubo2012}. The realization of this theoretical idea was experimentally verified in spinel MnSc$_2$S$_4$\cite{Gao2016, Gao2020}. The phase diagram of this material was studied for magnetic fields applied along the $[$100$]$ and $[$111$]$ crystal directions and the AF-SkL phase existed down to the lowest temperatures.\\ 
Our present study reveals that the MnSc$_2$S$_4$ antiferromagnet obeys the universal features of bulk skyrmion hosting materials and that magnetic anisotropy controls the stability of AF-SkL. Our neutron single crystal diffraction and Monte Carlo simulations demonstrate that for magnetic fields applied along the $[$1-10$]$ direction the triple-$\bf{q}$ phase vanishes and the conical phase occurs at very low temperatures which we could reach as in the experiments so in the simulations. 

\section{Recap on MnSc$_2$S$_4$}{\label{Sec2}}

\begin{figure*}[t]
\includegraphics[width=0.9\textwidth]{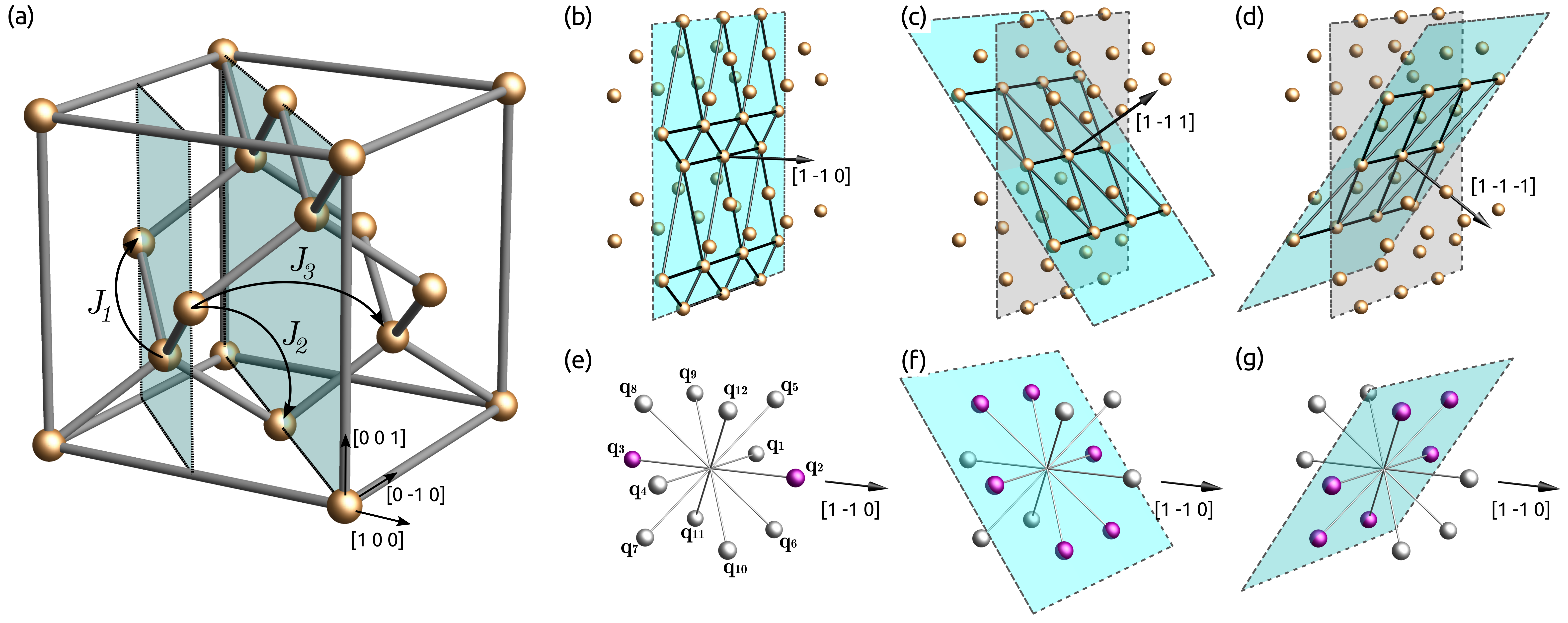}
\caption{(Color online) (a) Diamond lattice of the Mn$^{2+}$ ions consists of two interpenetrating face centered cubic ($fcc$) Bravais lattices. $J_1$, $J_2$ and $J_3$ indicate first, second and third neighbors exchange interactions, respectively. Panels (b)-(d) show the (1-10), (1-11) and (1-1-1) lattice planes. Panels (e-g) show the distribution for the arms of the propagation vector ${\bf{q}}=(\frac{3}{4}, \frac{3}{4}, 0)$ highlighting in purple (e) two arms ${\bf{q}_2}$ and ${\bf{q}_3}$, parallel to the magnetic field in [1-10], indicated by an arrow, (f) six arms corresponding to d$_4$- and (g) d$_1$- domains of the triple ${\bf{q}}$ phase.} 
\label{fig:DL}
\end{figure*}

In the cubic MnSc$_2$S$_4$ spinel the magnetic Mn$^{2+}$ ions (S=5/2) form a bipartite diamond lattice, with two face-centred cubic ($fcc$) sublattices A and B (Fig.~\ref{fig:DL}). This lattice can be frustrated when the second-neighbour $J_2$ coupling is antiferromagnetic ($J<0$, AF), as the Mn$^{2+}$ ions connected by the $J_2$ exchange build an AF ${fcc}$ sublattice which is highly frustrated. The ground state of the $J_1 - J_2$ model on the diamond lattice for the ferromagnetic (F) nearest-neighbour $J_1$ and strong AF $J_2$ couplings ($J_2$/$|J_1|>$1/8) is a manifold of massively degenerate spin spirals, named a spiral spin liquid\cite{Bergman2007}.

In MnSc$_2$S$_4$ strong frustration arising due to the large $J_2 /|J_1|$ ratio is released by a small third-neighbour $J_3$ coupling. In zero field the multi-step order sets in below $T_N$=2.3 K. The commensurate collinear state with the propagation vector ${\bf{q}}=(\frac{3}{4}, \frac{3}{4}, 0)$ transforms into the incommensurate short-living state with ${\bf{q}}_{icm}=(\frac{3}{4}\pm\delta, \frac{3}{4}\pm\delta, 0)$ and then back to the ground state commensurate helical state with the re-entrant ${\bf{q}}$. The propagation vector ${\bf{q}}=(\frac{3}{4}, \frac{3}{4}, 0)$ could have twelve arms listed in Table \ref{tab1}. 

Analysis of the intensity of the twelve $\langle \frac{3}{4}, \frac{3}{4}, 0 \rangle$ magnetic reflections unravels that in zero field the multi-domain single-${\bf{q}}$ helical (HL) structure exists forming twelve domains. In moderate magnetic fields a multi-domain triple-$\bf{q}$ phase occurs. One domain of this phase is described by three coplanar ${\bf{q}}_j$ vectors satisfying the rule $\sum_{j=1}^3{\bf{q}}_j = 0$ and 
four possible domains (d$_i$, $i$=1-4) are disclosed in Table~\ref{tab1} and Fig.~\ref{fig:DL}. 
Magnetic field along the $[$100$]$ crystal direction is equally inclined to all four domains, while the field along $[$111$]$ complies with the symmetry of only one domain. Exactly these domain distributions were observed in single crystal neutron diffraction experiments\cite{Gao2016, Gao2020}.\\

\begin{table} [h]
\caption{The twelve arms of the propagation vector ${\bf{q}}=(\frac{3}{4}, \frac{3}{4}, 0)$ with twelve domains of the helical or conical phases or four domains d$_{1-4}$ of the triple-$\bf{q}$ phase.
\label{tab1}}
\begin{ruledtabular}
\begin{tabular}{llllllll}
\textbf{q}$_1$=({\tq} {\tq} 0)&\textbf{q}$_7$=(-{\tq} 0 -{\tq})&\textbf{q}$_{12}$=(0 -{\tq} {\tq})&\textbf{q}$_4$&\textbf{q}$_5$&\textbf{q}$_{11}$&& d$_1$\\
\textbf{q}$_2$=({\tq} -{\tq} 0)&\textbf{q}$_8$=(-{\tq} 0 {\tq})&\textbf{q}$_{11}$=(0 {\tq} -{\tq})&\textbf{q}$_3$&\textbf{q}$_6$&\textbf{q}$_{12}$&& d$_2$\\
\textbf{q}$_3$=(-{\tq} {\tq} 0)&\textbf{q}$_5$=({\tq} 0 {\tq})&\textbf{q}$_{10}$=(0 -{\tq} -{\tq})&\textbf{q}$_2$&\textbf{q}$_7$&\textbf{q}$_9$&& d$_3$\\
\textbf{q}$_4$=(-{\tq} -{\tq} 0)&\textbf{q}$_6$=({\tq} 0 -{\tq})&\textbf{q}$_9$=(0 {\tq} {\tq})&\textbf{q}$_1$&\textbf{q}$_8$&\textbf{q}$_{10}$&&d$_4$\\
\end{tabular}
\end{ruledtabular}
\end{table}
These experimental observations could be explained by an effective spin Hamiltonian
\bea
\mathcal{H}&=&\sum_{ij}J_{ij}\Sp_i\cdot\Sp_j+3\,J_{||}\sum_{ij\in NN}\left(\Sp_i\cdot \hat{r}_{ij} \right)\left(\Sp_j\cdot \hat{r}_{ij} \right)\nonumber\\
&&+A_4\sum_{i,\alpha=x,y,z}(S^{\alpha}_i)^4-g\mu_B\,{\bf B}\sum_i\Sp_i
\label{eq:Hamiltonian}
\eea
where $J_{ij}$ is the exchange coupling (up to the third neighbours) between the Heisenberg spins $\Sp_i$ and $\Sp_j$. $J_{||}$ is the anisotropic coupling between the nearest-neighbours connected by the unitary vectors $\hat{r}_{ij}$  along the $\langle$110$\rangle$ axes, $A_4$ is fourth-order single-ion anisotropy constant. The last term in Eq. (\ref{eq:Hamiltonian}) is the Zeeman coupling between the spins and a magnetic field ${\bf B}$; $g$ and $\mu_B$ are the Land\'e g-factor and the Bohr magneton. 

From combined inelastic neutron scattering and Monte-Carlo simulations study\cite{Gao2020} a set of exchange parameters $J_1 = -0.31$K, $J_2 = 0.46$K, $J_3 = 0.087$K, $J_{||}=-0.01$K and $A_4=0.0016$K was extracted. It reproduces the zero field helical ground state, the spin-wave excitations and the (T, B) phase diagram for magnetic fields applied along the $[$100$]$ and $[$111$]$ crystal directions.\\ 
Moreover, the triple-$\bf{q}$ phase with the finite scalar spin chirality could be rationalized as a fractional antiferromagnetic skyrmion lattice and the analytical expression for this AF-SkL was found.

\section{Monte Carlo simulations}{\label{SecMC}}

The present study aims to explore the stability of the fractional AF-SkL lattice in MnSc$_2$S$_4$ applying field along the [1$\overline{1}$0] direction. 
The classical spin model of Eq. (\ref{eq:Hamiltonian}) and the set of $J_{1,3}$-$J_{||}$-$A_4$ parameters as reported in Ref.~[\onlinecite{Gao2020}] were implemented on a $N=8\times L^3$ ($L=8$) lattice with periodic boundary conditions. In this calculation, the spin lattice has been cooled slowly to low temperatures and the averaging over $100$ independent attempts has been performed. In order to identify different phases and to compare with neutron diffraction results, we computed the magnetic structure factor defined as
\begin{equation}
F(\mathbf{q})= \frac{f(\mathbf{q})^2}{N} 
\sum_{ij} \langle\Sp_i\cdot\Sp_j\rangle e^{i \mathbf{q}\cdot \hat{r}_{ij}}
\label{eq:F}
\end{equation}
and the measured quantity $F_{\perp {\bf q}}^2({\bf q})$ where $\perp {\bf q}$ indicates that the spin-spin correlation only includes spin components perpendicular to ${\bf q}$. 
In Eq. (\ref{eq:F}) $\noindent \langle\Sp_i\cdot\Sp_j\rangle$ are the correlations between the magnetic moments at the sites $i$ and $j$, $N$ is the number of magnetic ions and $f(\mathbf{q})$ is the Mn$^{2+}$ magnetic form factor.

\begin{figure}
\includegraphics[width=0.8\columnwidth]{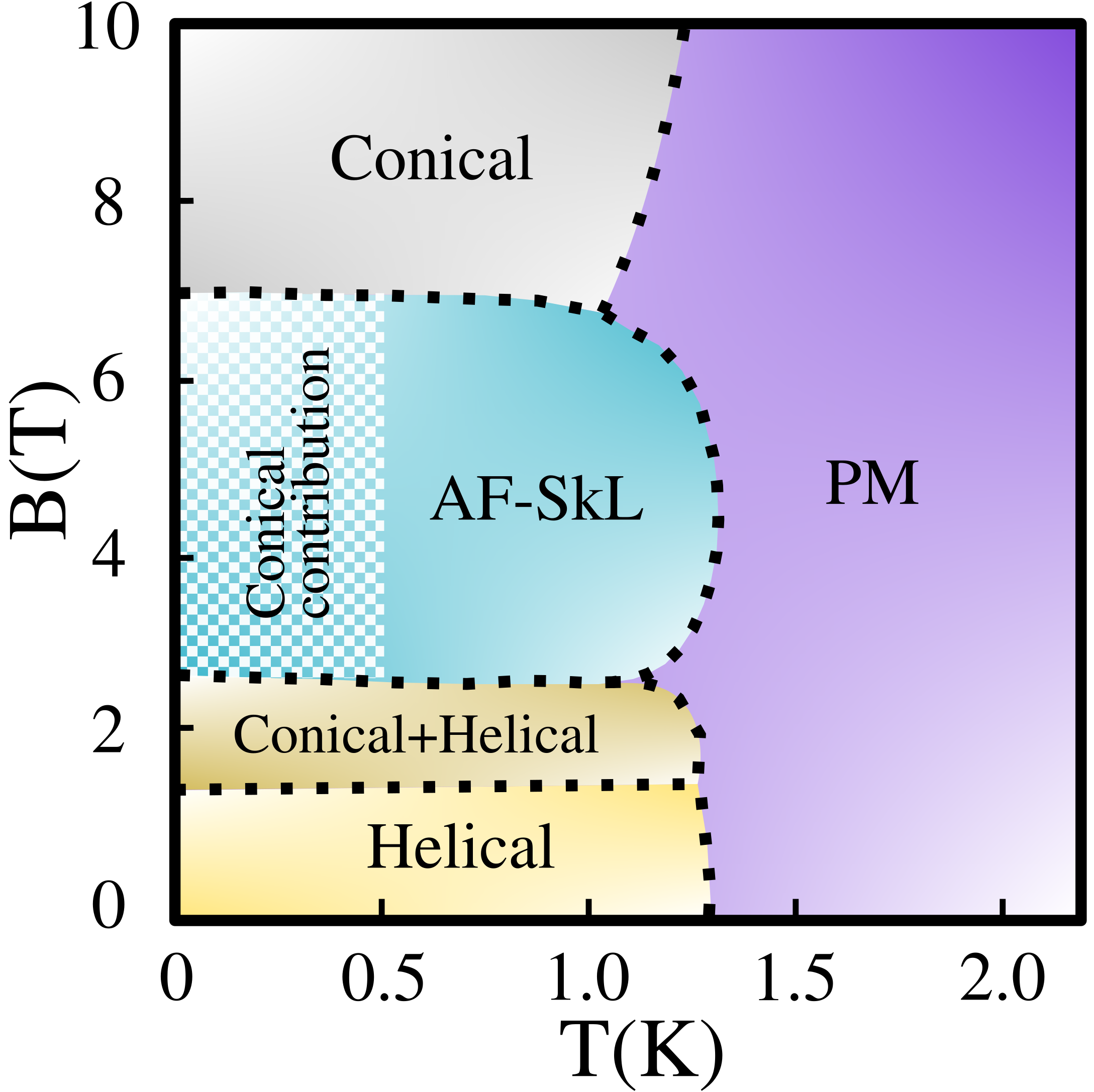}                     
\caption{(Color online) Magnetic phase diagram of the model in Eq. (\ref{eq:Hamiltonian}) as a function of the temperature T(K) and external magnetic field B(T) along the [1$\overline{1}$0] direction.} 
\label{fig:PD-MC}
\end{figure}

Fig.~\ref{fig:PD-MC} presents the theoretical (T, B$_{1-10}$) phase diagram with three ordered phases - helical (HL), antiferromagnetic skyrmion lattice (AF-SkL) and conical (CN). At zero and low fields ($B\leq 1.2$T) only the helical phase exists. Then, on field growth a strong competition between the HL and CN phases occurs. As field increases further to $B$=3.5 T the AF-SkL phase emerges. At lowest temperatures there is an additional non-zero contribution from the conical phase in the predominantly AF-SkL region. The CN phase is favored at higher magnetic fields, where the AF-SkL phase vanishes. The boundaries of these and other phases were determined from the intensity distribution of the $\langle \frac{3}{4}, \frac{3}{4}, 0 \rangle$ magnetic peaks, as we will discuss in detail in the following subsections.

\begin{figure}
\includegraphics[width=0.8\columnwidth]{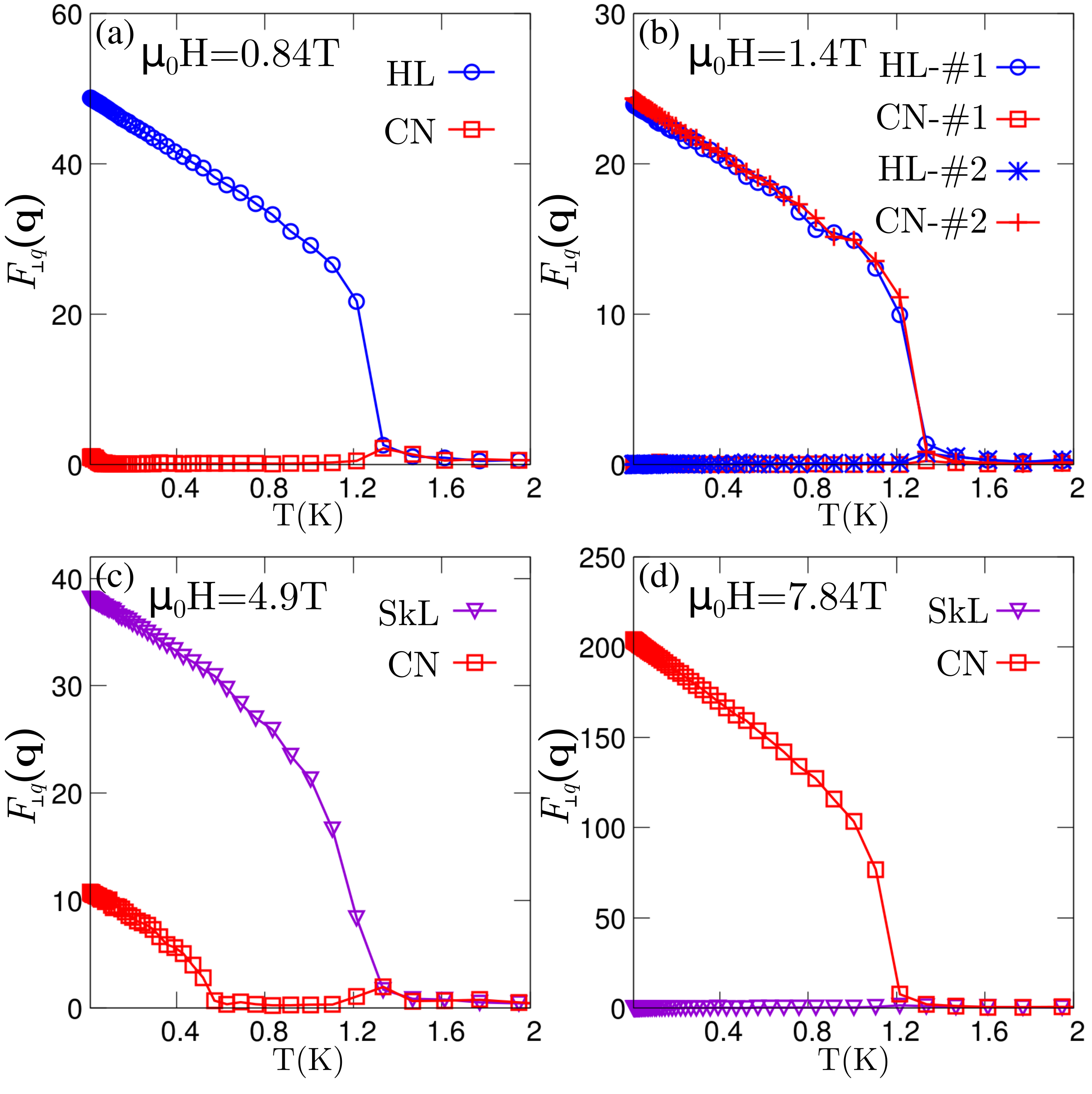}
\caption{Neutron intensity as a function of temperature and magnetic field along [1$\overline{1}$0] averaged over 100 MC runs. (a) Contributions  at $B=0.84$T averaged over ten  single-${\bf{q}}$ domains of the helical structure (blue) and over two single-${\bf{q}}_n$ domains of the conical structure (red). (b) Intensities obtained in two independent MC runs ($\#$1, $\#$2) at $B=1.4$T. (c) Contributions for two d$_{1,4}$ domains of triple-${\bf{q}}$ denoted SkL (violete) and two CN domains (red) at $B=4.9$T. (d) The same intensities for SkL and CN at high fields, $B=7.84$T.
} 
\label{fig:Sq-T}
\end{figure}
%

\subsubsection{Helical phase}

\begin{figure}[htb]
\includegraphics[width=1.0\columnwidth]{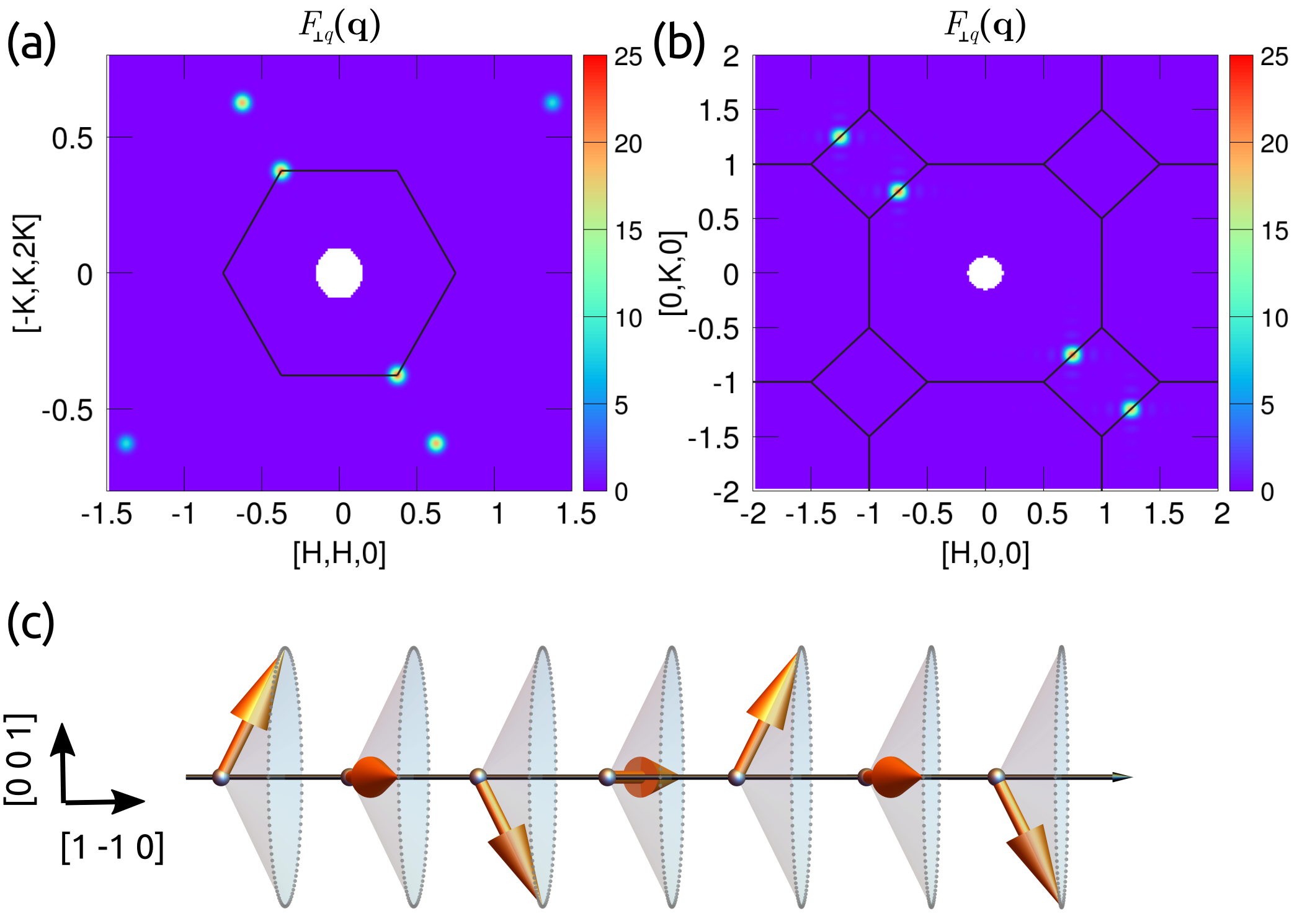}
\caption{(Color online) 
(a) A typical intensity distribution after one individual run at $T=0.4$K and $B=0.84$T. The magnetic intensity corresponds to ${\bf{q}}_6=(\frac{3}{4},0,-\frac{3}{4})$ and belongs to the helical phase.
(b) Same at $B=1.4$T, the magnetic intensity appears at ${\bf{q}}_2=(\frac{3}{4}, -\frac{3}{4}, 0)$ and this MC run results in the conical phase. 
(c) The spin arrangement in the conical phase.} 
\label{fig:Sq-MC}
\end{figure}

In low magnetic fields ($B< 1.2$T) our MC runs result in finite intensity in the (1-11) and (1-1-1) reciprocal lattice planes (Fig.~\ref{fig:DL} (f), (g)). These planes comprise ten arms of the propagation vector ${\bf{q}}=(\frac{3}{4}, \frac{3}{4}, 0)$, which are listed in Table \ref{tab1} (top and bottom rows). 
The intensity distribution for all MC runs and respective magnetic arrangements imply realisation of the helical phase.
One example of such MC $F_{\perp {\bf q}}({\bf q})$ map is outlined in Fig.~\ref{fig:Sq-MC}(a), it corresponds to the ${\bf{q}}_6$ domain. Within a significant (T, B) region no contributions from the remaining ${\bf{q}_{2}}$, ${\bf{q}_{3}}$ arms emerge (see Fig.~\ref{fig:Sq-T}(a)). 

At higher magnetic fields ($B\gtrsim 1.2$T) the helical and conical phases compete and the reflections of both phases appear. In our MC runs for $B=1.4$T either the HL phase or the CN phase with propagation vector  ${\bf{q}}_2=(\frac{3}{4}, -\frac{3}{4}, 0)$ emerge (see Fig.~\ref{fig:Sq-T}(b)). 
Fig.~\ref{fig:Sq-MC}(b) presents a $F_{\perp {\bf q}}^2({\bf q})$ map for the MC run converging to the conical phase with ${\bf{q}}_2$.

\begin{figure}
\includegraphics[width=0.9\columnwidth]{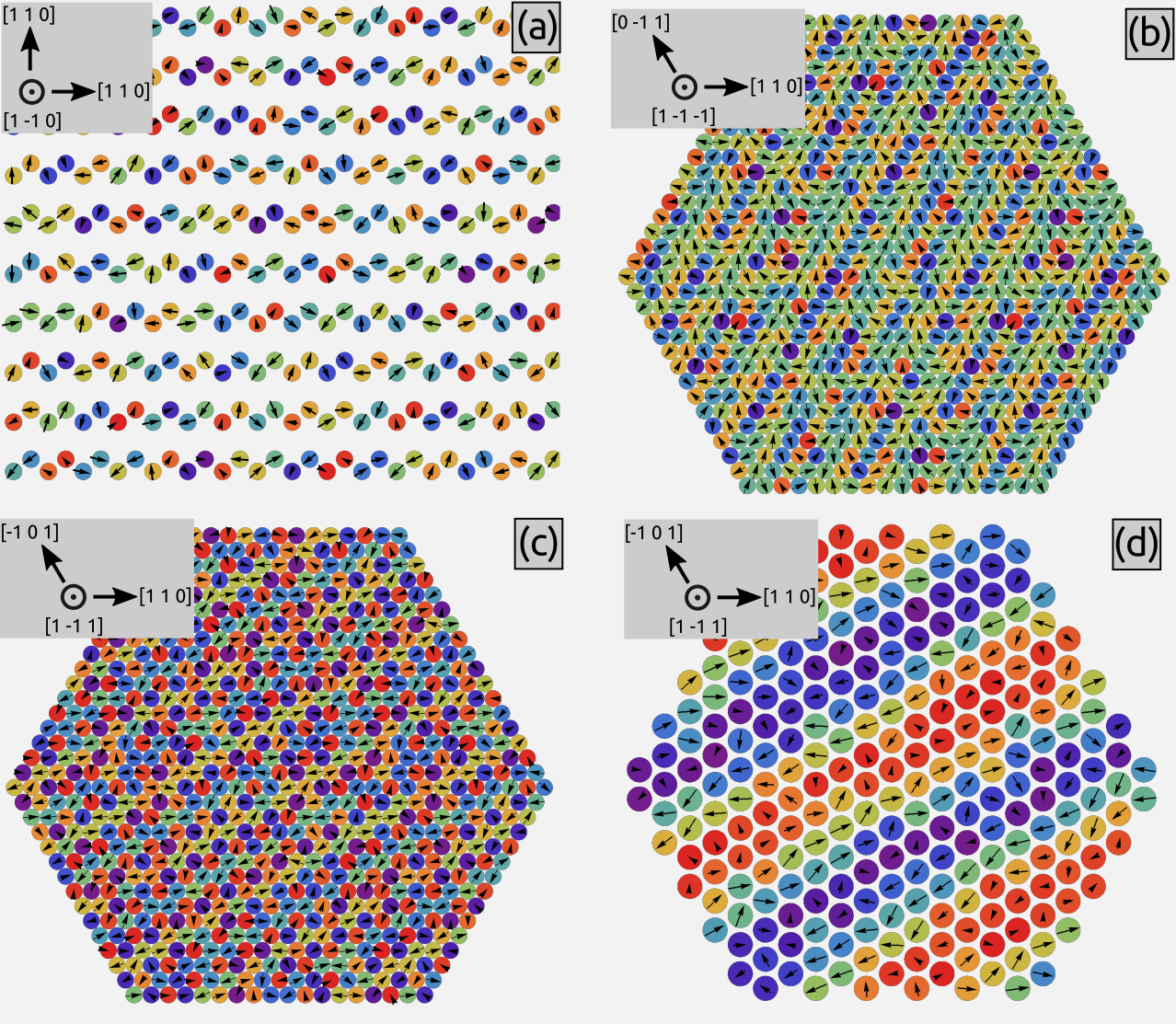}
\caption{(Color online) Helical texture at $B=0.56$T and $T=0.92$K in different planes. (a): (1-10), (b): (1-1-1), (c): (1-11); (d) spin texture in one sublattice of the triangular layer in the (1-11) planes. The color scheme represents the z-component of the magnetic moment in the local coordinate system of each panel with S=-1 in blue, S=1 in red. The arrows shown the in-plane spin component.}
\label{fig:H-texture}
\end{figure}

To characterize the single-$\bf{q}$ helical structure we explore the magnetic arrangement in real space. A typical (1-10) plane obtained from one MC run at $B=0.56$T and $T=0.92$K is shown in Fig.~\ref{fig:H-texture}(a). 
No ordered arrangement is evident. 
So we inspect separately the (1-11) and (1-1-1) planes. 
While no clear regularity is seen in (1-1-1) (Fig. \ref{fig:H-texture}(b)),
a distinct ordered structure is apparent within the (1-11) plane (Fig.~\ref{fig:H-texture}(c)). Further on, when subdividing this triangular layer into three sublattices (Fig.~\ref{fig:H-texture}(d)) 
(that is twice depleting the layer as in Ref.~[\onlinecite{Gao2020}]) 
the superposition of three interpenetrating helices is obvious.
Thus the MC run presented in Figs. \ref{fig:H-texture} (c, d) displays one of the possible single-$\bf{q}$ associated with the helical structure.
The symmetry between the (1-1-1) and (1-11) planes is broken. 
In this example, only  one domain of the helical structure with spin-correlations in the (1-11) plane arises, while domains with the HL long-range arrangement in the (1-1-1) plane are absent.
This spontaneous symmetry breaking is confirmed by other MC runs, where the ordered structure is seen in either (1-11) or (1-1-1) planes.

The situation is analogous to the observations in the triangular\cite{Rosales2015,Osorio2017} and kagome\cite{Villalba2019} antiferromagnetic lattices, where the underlying structures are revealed upon splitting the full arrangement into simpler sublattices.
%

\subsubsection{Conical phase}

Further increase of magnetic field along [1$\overline{1}$0] favors the single-$\bf{q}$ conical phase with the propagation vector along the field (Fig.~\ref{fig:DL}(e)).
The stability of the CN phase in the (T,B) phase diagram could be monitored following the intensity of the ($\frac{3}{4}, -\frac{3}{4}, 0$) reflection. Fig. ~\ref{fig:Sq-T}(d) shows that $\bf{q}_2$ clearly dominates other $\bf{q}$-arms at field as high as $B=7.84$T.
In fact, no other arm is present in this region of the phase diagram (Fig.~\ref{fig:Sq-MC}(b)), thus this is a single-$\bf{q}$ structure.\\
The corresponding spin texture in real space is sketched in Fig.~\ref{fig:Sq-MC}(c). 
It is composed of chains of the Mn moments propagating along the [1,-1,0] directions with a helical component orthogonal to ${\bf{q}}$ and a ferromagnetic component along ${\bf{q}}$.

\subsubsection{Fractional AF-SkL phase}
Monte Carlo simulations of the model proposed in Eq. (\ref{eq:Hamiltonian}) stabilize the fractional AF-SkL for magnetic fields along [111]\cite{Gao2020}. This direction is perpendicular to the triangular layers of the spinel structure, where the spin texture develops. Here we show that this AF-SkL phase emerges also when magnetic field is applied along [1-10]. Remarkably, the skyrmion texture is not perpendicular to the external field, but spans in the (1-11) and (1-1-1) planes, i.e. in the triangular layers. 

\begin{figure}
\includegraphics[width=1.0\columnwidth]{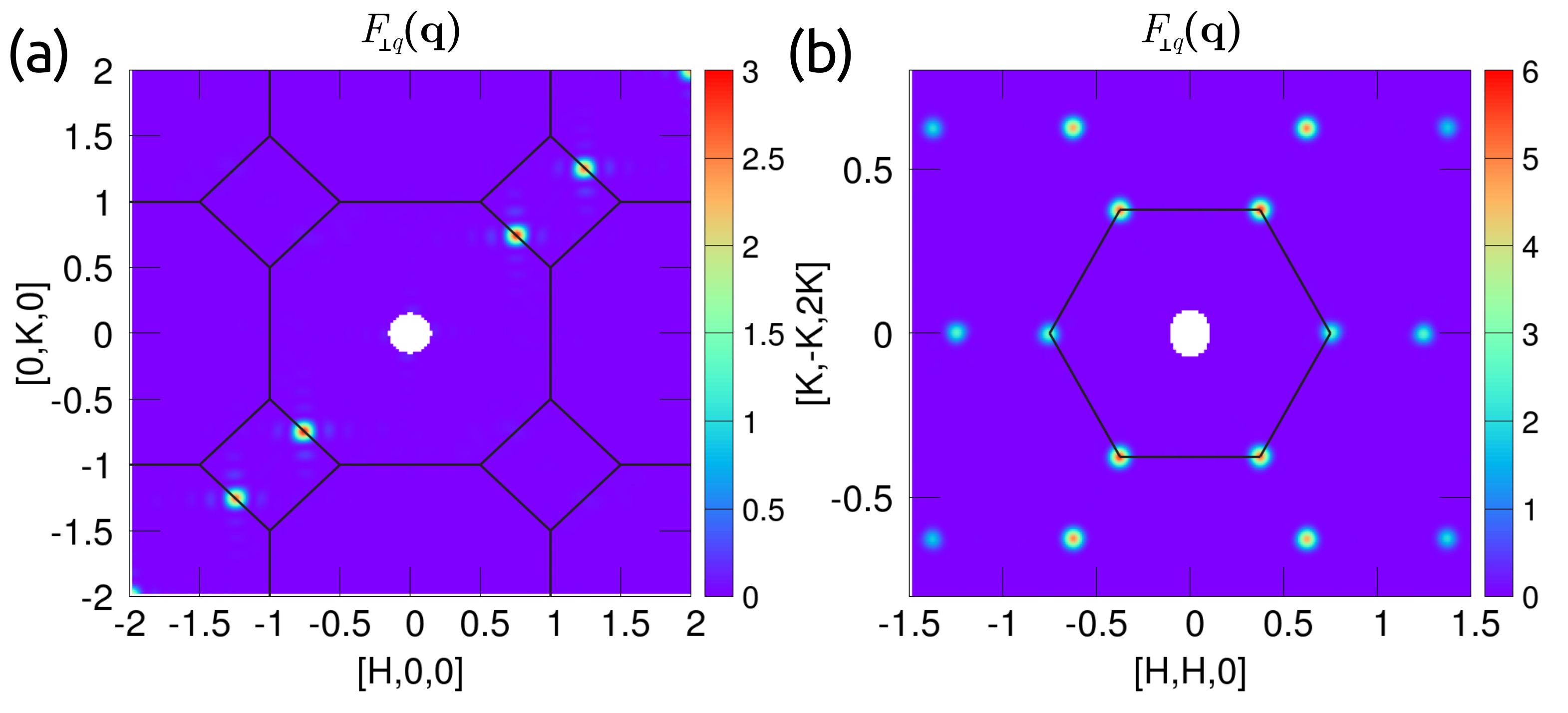}
\caption{(Color online) Magnetic intensities obtained by MC simulations at  $T=0.96$K and $B=4.9$T in two different planes of the reciprocal lattice. This distribution corresponds to the AF-SkL phase.} 
\label{fig:Sq-SkL}
\end{figure}

The emergence of the AF-SkL phase could be identified inspecting intensity distribution of the twelve $\langle\frac{3}{4}, \frac{3}{4}, 0\rangle$ reflections. 
These reflections are grouped into contributions of four domains of the triple-$\bf{q}$ listed in Table \ref{tab1}. Fig.~\ref{fig:Sq-T} (c) presents temperature dependence of intensity from MC runs at $B=4.9$T. Two separate groups are apparent - two ($\pm \bf{q}_2$) and the rest ten.
Such T-dependence implies that these groups arise from the CN and AF-SkL phases, respectively.
The ten peaks belonging to the d$_1$, d$_4$ AF-SkL domains clearly dominate till T$_N$=1.3K. However, at low temperatures ($T \lesssim 0.5K$) the $\pm \bf{q}_2$ intensity is non-zero, implying coexistence of AF-SkL with the conical phase. We present magnetic intensities for the pure AF-SkL phase at $T=0.96K$ and $B=4.9$T in Fig. \ref{fig:Sq-SkL} (a,b). Here the intensity distribution typical for the d$_1$-domain is apparent.

\begin{figure}
\includegraphics[width=0.9\columnwidth]{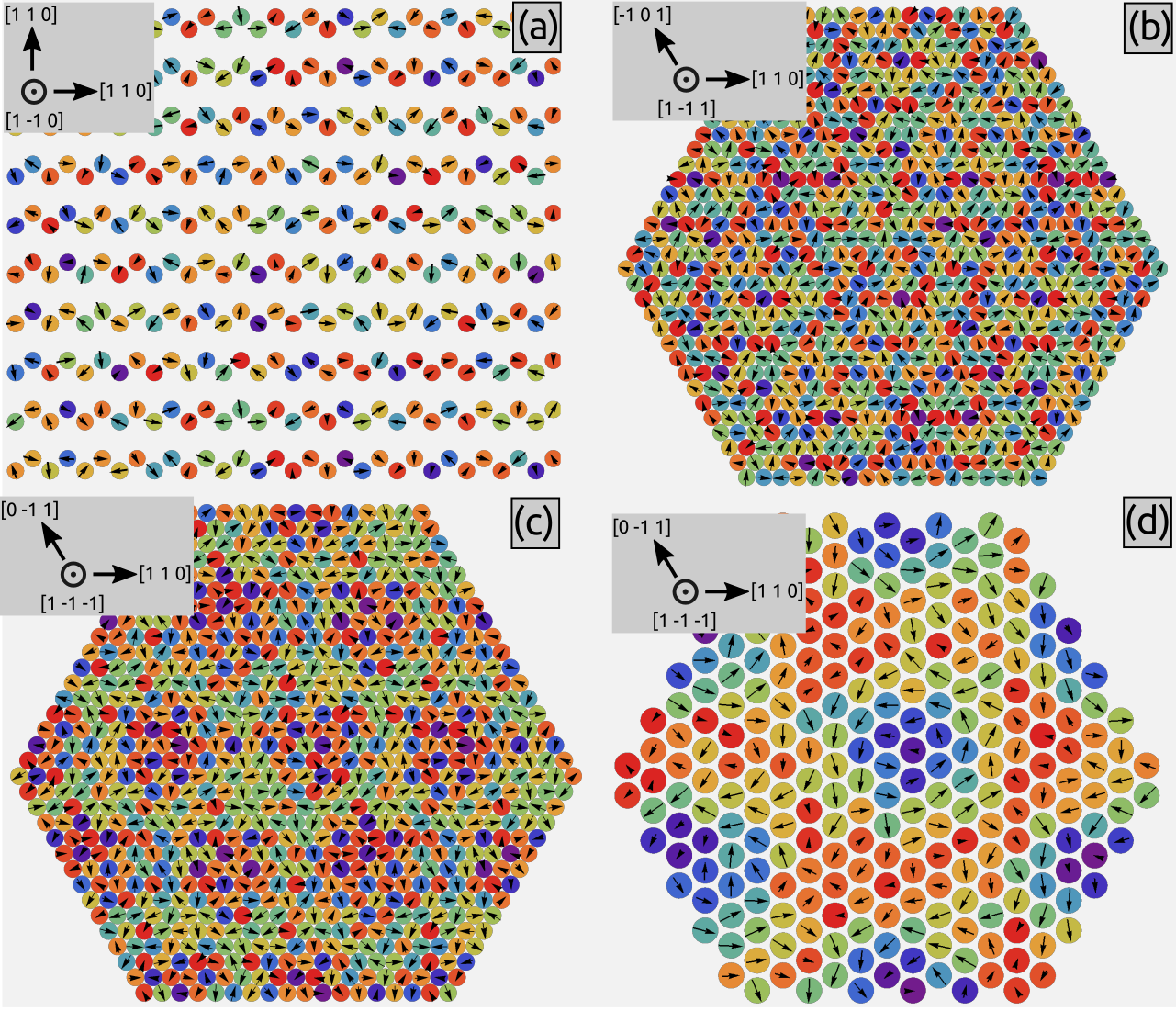}
\caption{(Color online) Magnetic texture in the triple-$\bf{q}$ phase at $B=4.9$T and $T=0.96$K in different planes. (a): (1-10), (b): (1-11), (c): (1-1-1); (d) spin texture in one sublattice of the triangular layer in the (1-1-1) planes. The color scheme is the same as in Fig. \ref{fig:H-texture}.
} 
\label{fig:SkL-texture}
\end{figure}

\begin{figure}
\includegraphics[width=0.8\columnwidth]{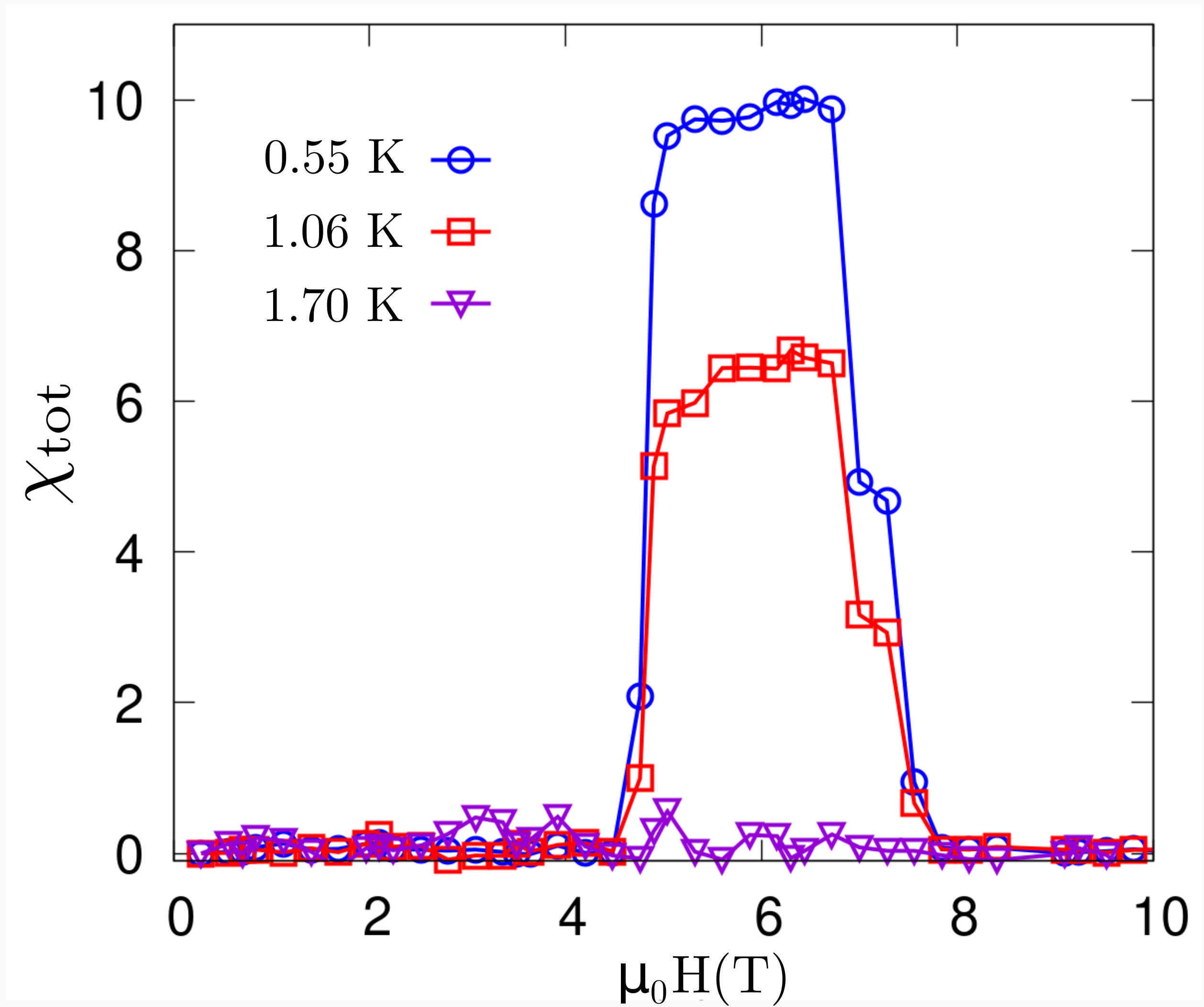}
\caption{(Color online) Total scalar spin chirality computed for selected MC final configurations at three representative temperatures.
} 
\label{fig:ChiQvsB}
\end{figure}

Now we look at the corresponding spin texture in real space. In Fig.~\ref{fig:SkL-texture} a typical arrangement at $T=0.96$K and  $B=4.9$T in three different planes: (1-10), (1-1-1) and (1-11) is shown.
Like in the helical phase, the true order cannot be identified easily in the (1-10) planes (see Fig.~\ref{fig:SkL-texture}(a)). However, if we inspect the (1-11) and (1-1-1) planes, the famous fractional antiferromagnetic skyrmion lattice \cite{Gao2020} is visible (see Figs.~\ref{fig:SkL-texture}(c) and \ref{fig:SkL-texture}(d)).
Again, like in the helical phase, different numerical simulations show that the SkL pattern is stabilized either in the (1-11) plane or in the (1-1-1) plane, never in both planes simultaneously. This means that either the d$_1$- or d$_4$- domain of the AF-SkL phase emerges.
To further study the AF-SkL phase, we compute the absolute value of the total scalar spin chirality (discrete topological charge) defined as
\begin{equation}
\chi_{\text{tot}}=\langle\frac{1}{8\pi}\sum_{\triangle}\Sp_{i}\cdot(\Sp_{j}\times\Sp_k)\rangle
\end{equation}
where $\triangle$ indexes all the elementary triangles of the sites $i$, $j$ and $k$ in the (1-11) and (1-1-1) layers. A finite $\chi_{\text{tot}}$ is expected for topological textures, whereas it would vanish for the helical and conical phases. In Fig.~\ref{fig:ChiQvsB} $\chi_{\text{tot}}$ as a function of magnetic field is presented for three representative temperatures. AF-SkL is clearly identified at 0.55K and 1.06K, but not at 1.70K.
%

\section{Experiment}
\begin{figure}
\includegraphics[width=0.9\columnwidth]{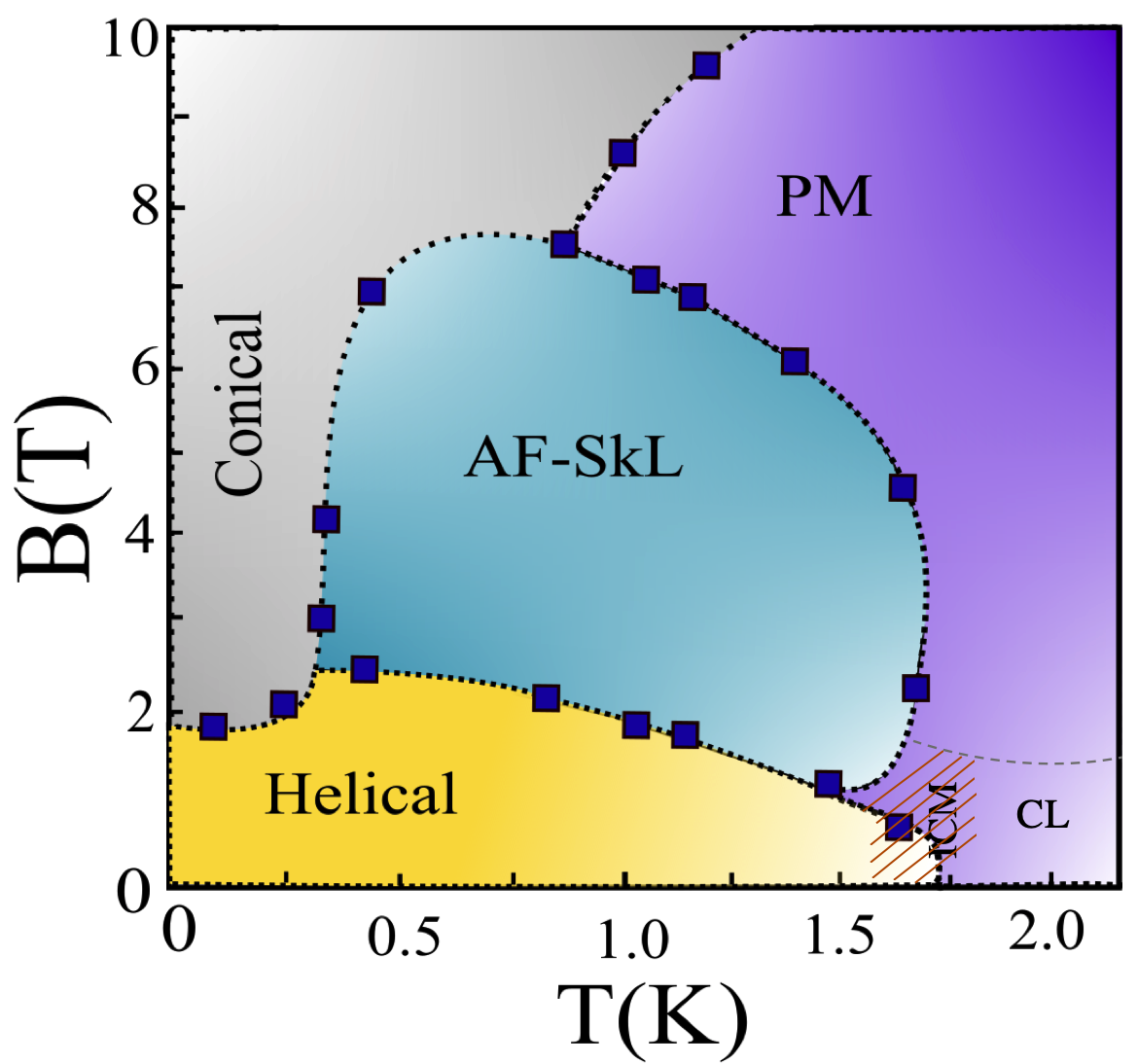}
\caption{(Color online) Magnetic phase diagram obtained by single crystal neutron diffraction with magnetic field B[T] along the [1$\overline{1}$0] direction.} 
\label{fig:PD_exp}
\end{figure}

\begin{figure}
\includegraphics[width=0.8\columnwidth]{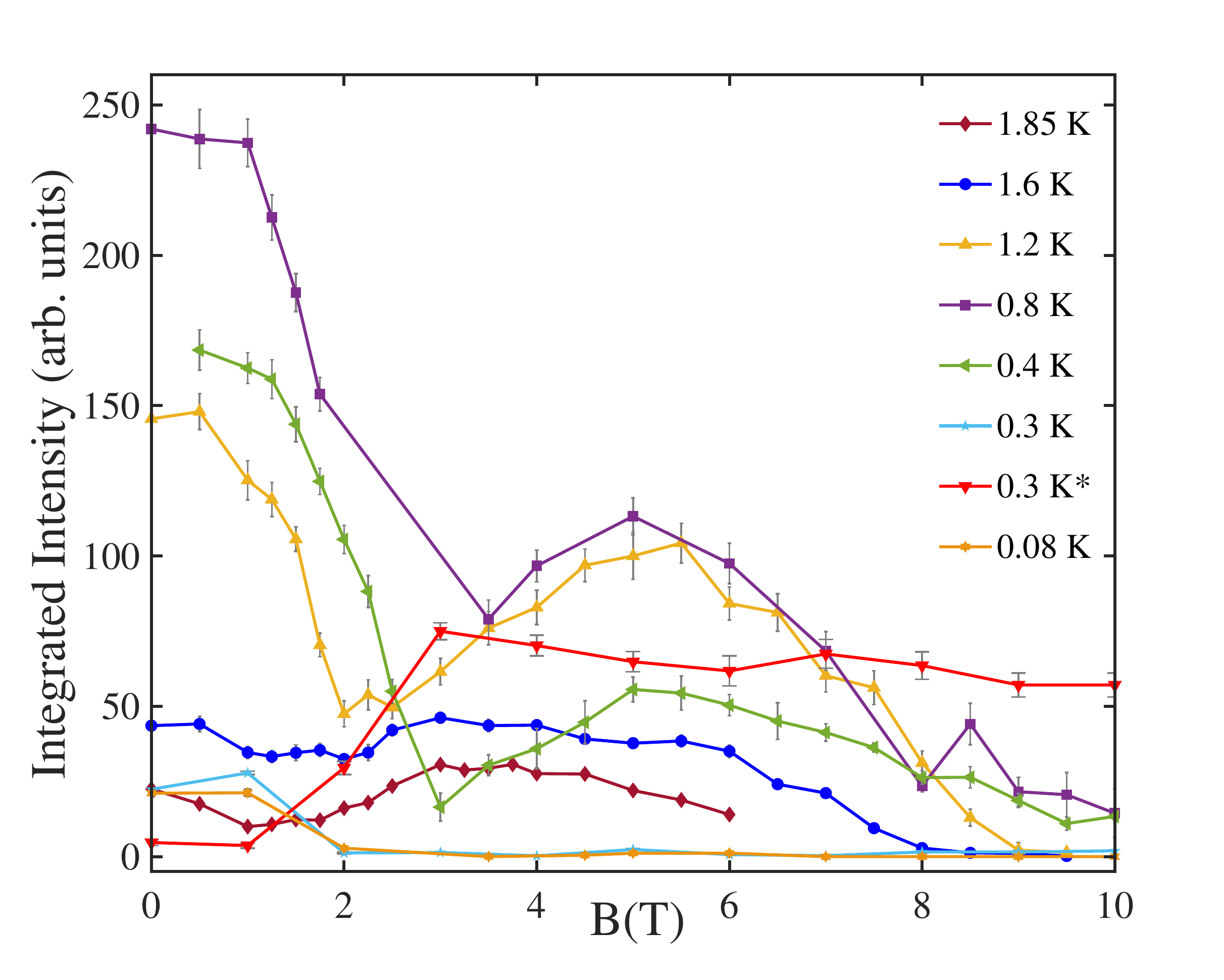}
\includegraphics[width=0.8\columnwidth]{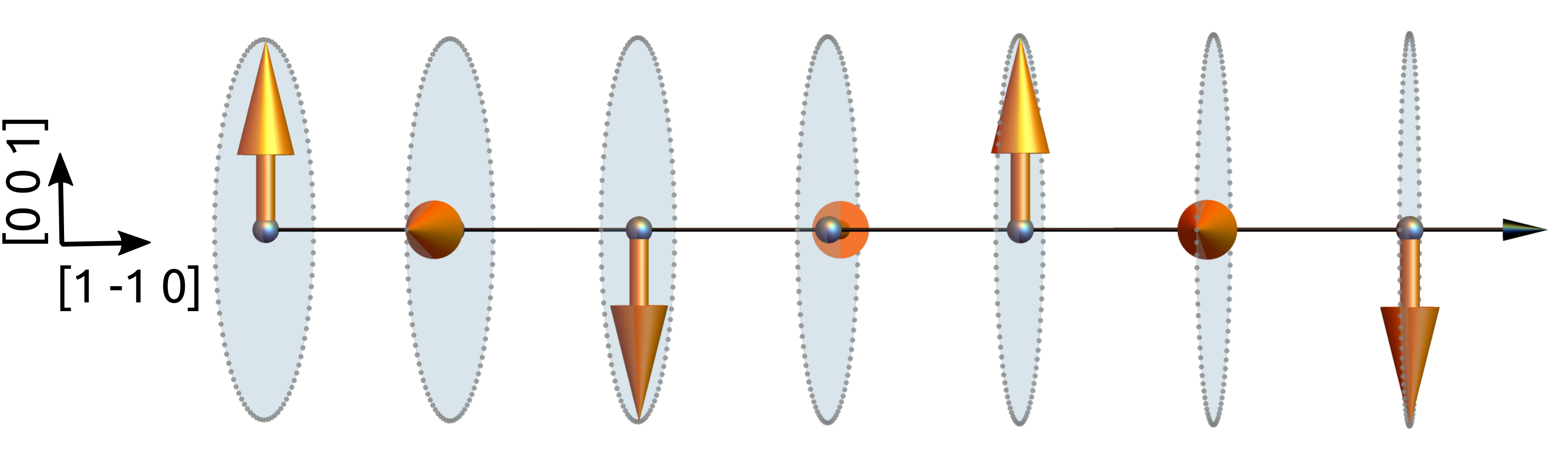}
\caption{(Color online) (a) Field and temperature evolution of magnetic intensity at the ($\frac{3}{4}, \frac{3}{4}, 0$) position. Intensity for the ($\frac{1}{4}, -\frac{1}{4}, 1$) peak measured at T=0.3 K is presented in red and marked by star. (b) Helical part of the cone structure from single crystal neutron diffraction at T=0.1 K and B=2.5 T.} 
\label{fig:CN_exp}
\end{figure}

{\label{Sec3}}
Neutron single crystal diffraction (ND) was used to generate the experimental (T,B) phase diagram and to verify the structure of the conical phase. The single crystals grown with the chemical transport reaction technique and used in previous diffraction experiments \cite{Gao2016,Gao2020} were utilized. This time the experiments were performed on the neutron single-crystal diffractometer Zebra at the Swiss Spallation Neutron Source, SINQ, Villigen and on the D23 diffractometer at the Institut Laue-Langevin (ILL), Grenoble using the normal beam geometry, dilution inserts and vertical magnets. The crystals
were oriented with the [1-10] crystal axis vertical, along the applied magnetic field. The neutron wavelengths $\lambda$=2.314 \AA~and 1.383 \AA~ selected by the PG(200) and Ge(220) monochromators and a 10 T magnet were used on Zebra. The wavelength 1.273 \AA~ from a Cu(200) monochromator and a 12 T magnet were used on D23.\\
The phase diagram constructed from our experiments is presented in Fig. \ref{fig:PD_exp} (a). The boundaries of different phases were determined following the intensity of the ${\bf{q}_1}=(\frac{3}{4}, \frac{3}{4}, 0$) magnetic reflection (Fig. \ref{fig:CN_exp} (a)).
When the intensity was changing abruptly, increasing, as in the AF-SkL region, or vanishing, as in the CN phase, we measured intensity of other ${\bf{q}}$ arms to identify which magnetic phase was entered. As the $(\frac{3}{4}, -\frac{3}{4}, 0$) reflection could not be reached due to the limits of the  experimental setup, the ($\frac{1}{4}, -\frac{1}{4}, 1$) reflection of the ${\bf{q}_2}$ arm was measured instead (Fig. \ref{fig:CN_exp} (b)).\\
In order to determine the antiferomagnetic component of the
conical structure a set of magnetic reflections of the ${\bf{q}_2}=(\frac{3}{4}, -\frac{3}{4}, 0)$ arm was measured at $T=0.1$K and $B=2.5$T. The helical component of the conical structure proposed from the MC runs gives good fit to the data ($R_{F^2}$=7.3 \%, $\chi^2=2.16$ for 31 measured reflections). The two helices with the origin at the sublattices A and B are circular, with almost equal components in the $[1,1,0]$- plane ($M_{[1,1,0] }$=4.65(9)) $\mu_B$) and out-of-plane ($M_z$=4.7(2) $\mu_B$).
In the AF-SkL region the magnetic reflections corresponding to two domains of the AF-SkL phase, d$_1$ and d$_4$, were detected.

\section{Discussion and Summary}{\label{Sec4}}

Our study brings new insights into the (T, B) phase diagram of the antiferromagnetic spinel MnSc$_2$S$_4$ and SkL hosting materials in general.\\
Firstly, for the B//[1-10] field direction the conical phase was found for MnSc$_2$S$_4$. This phase was not detected for the other principal field directions\cite{Gao2016,Gao2020}. In our MC runs the CN phase coexists with the helical phase at low magnetic fields and with the AF-SkL phase at higher fields. In the ND experiments the HL-CN transition is more perceptive and a single-phase region instead of two-phase regions is observed. We suspect that the mixed regions in the MC simulations is a consequence of freezing of the adjacent phases at these low temperatures.
The conical phase has the helical component orthogonal to the propagation direction from both, ND experiment and MC runs, while the ferromagnetic component is along the propagation direction according to MC.\\
Secondly, our MC simulations show that the AF-SkL phase for MnSc$_2$S$_4$ with B//[1-10] is the same as the one found for the [111] field direction\cite{Gao2020}. The SkL tubes propagate along the [1-11] or [1-1-1] axes and the spin texture of the Bloch-type develops in the triangular layers perpendicular to these directions. 
The texture is a linear combination of helices. As nearest spins within the triangular layers are almost anti-parallel, we select next-nearest, almost parallel spins, which form a depleted triangular sublattice to visualize the swirls in Fig.~\ref{fig:H-texture} (d). The distance between their centers is $\approx$ 55 \AA\, and they can be viewed as pairs of incipient meron and antimeron \cite{Gao2020}. The scalar spin chirality is finite and non-integer in this topological texture.\\

The orientation of skyrmions in MnSc$_2$S$_4$ is locked to the anisotropy axis defined by the $J_{||}$ anisotropic exchange and fourth-order single-ion anisotropy $A_4$.
The relevance of the $J_{||}$ interaction in the formation of the fractional AF skyrmion lattice was discussed recently in Ref.[\onlinecite{Gao2020}] (see Supp. Material), while the single-ion anisotropy is a key factor stabilizing skyrmion textures in frustrated magnets \cite{Leonov2015}, since it lowers the energy of SkL with respect to the other phases at moderate magnetic fields.
Skyrmions do not rotate following the applied magnetic field in the MnSc$_2$S$_4$ host as it happens, for example, in the soft B20 Bloch-type SkL hosts\cite{Muelbauer2009,Seki2012}.
In our MC runs the domains equally inclined to the applied magnetic field get stabilised. In this aspect MnSc$_2$S$_4$ is closer to the family of lacunar spinels GaV$_4$X$_8$ (X=S, Se),\cite{Gross2020} where the Neel-type skyrmions are locked to the polar axis. For this last case rotation of the applied field
modifies domain walls where distortion of the skyrmions and displacement of their cores takes place.\cite{Geirhos2020}.\\
Furthermore the case of MnSc$_2$S$_4$ allows to generalize that the magnetic anisotropy is often a decisive factor for stabilization of the SkL phases. The nature of the dominant coupling: ferromagnetic, antiferromagnetic or RKKY, as well as, is the type of the texture: Bloch- or Neel-, make each material hosting skyrmions unique defining the multi-parameter space for the existence of SkL lattices. Yet, the anisotropy in any feasible form - single-ion anisotropy, anisotropy of exchange, dipolar interaction - defines the exact region of stability for the SkL-phase in the (T, B) phase diagram. 

{\it Acknowledgments.} 
The neutron diffraction experiments were performed at SINQ, Paul Scherrer Institute, Villigen, Switzerland and at ILL, Grenoble, France. This work was supported by the Swiss National Science Foundation (Grant No. 200020-182536). 
 H. D. R. and F. A. G. A. are partially supported
by CONICET (PIP 2021-11220200101480CO), and SECyT UNLP PI+D X792, X788 and X893. F. A. G. A.
acknowledges support from PICT 2018-02968. V. T. and L. D.were partly supported by the Deutsche Forschungsgemeinschaft (DFG) through Transregional Research Collaboration TRR 80 (Augsburg, Munich, and Stuttgart), and by the project ANCD 20.80009.5007.19 (Moldova)

\clearpage
\onecolumngrid

\bibliographystyle{apsrev4-1}

\end{document}